\documentclass[aps,prl,preprint,groupedaddress]{revtex4-2}
\usepackage{graphics, epsfig,colordvi}
\usepackage{kantlipsum,setspace}
\def\bx{{\bf x}}

\def\br{{\bf r}}
\def\brp{{\bf r'}}

\begin{document}
\baselineskip = 15 pt

% Use the \preprint command to place your local institutional report
% number in the upper righthand corner of the title page in preprint mode.
% Multiple \preprint commands are allowed.
% Use the 'preprintnumbers' class option to override journal defaults
% to display numbers if necessary
%\preprint{}

%Title of paper
\title{Vision: looking and seeing through our brain's information bottleneck}

% repeat the \author .. \affiliation  etc. as needed
% \email, \thanks, \homepage, \altaffiliation all apply to the current
% author. Explanatory text should go in the []'s, actual e-mail
% address or url should go in the {}'s for \email and \homepage.
% Please use the appropriate macro foreach each type of information

% \affiliation command applies to all authors since the last
% \affiliation command. The \affiliation command should follow the
% other information
% \affiliation can be followed by \email, \homepage, \thanks as well.
\author{Li Zhaoping}
\email[]{li.zhaoping@tuebingen.mpg.de}
%\homepage[]{Your web page}
%\thanks{}
%\altaffiliation{}
\affiliation{Max Planck Institute for Biological Cybernetics, University of T\"ubingen, Germany}

%Collaboration name if desired (requires use of superscriptaddress
%option in \documentclass). \noaffiliation is required (may also be
%used with the \author command).
%\collaboration can be followed by \email, \homepage, \thanks as well.
%\collaboration{}
%\noaffiliation

\date{\today}

\begin{abstract} % typically limited to 250 words according to ChatGPT, currently 293 words
Our brain recognizes only a tiny fraction of sensory input,
due to an information processing bottleneck. This blinds us
to most visual inputs. Since we are blind to this
blindness, only a recent framework highlights this bottleneck by
formulating vision as mainly looking and seeing.
Looking selects a tiny fraction of visual information for
progression through the bottleneck, mainly
by shifting gaze to center an attentional spotlight.
Seeing decodes, i.e., recognizes, objects within the selected information.
Since looking often occurs before seeing and  evokes limited awareness, 
humans have the impression of seeing whole scenes clearly.
According to the new framework, the bottleneck 
 starts from the output of the primary visual cortex (V1) to downstream 
brain areas. This is motivated by the evidence-backed V1 Saliency Hypothesis (V1SH)
that V1 creates a saliency map of the visual field to guide looking.
Massive visual information loss downstream from V1 makes
seeing  vulnerable to ambiguity and illusions (errors). 
To overcome this, feedback from downstream to upstream areas 
such as V1 queries for additional relevant information.
An integral part of this framework is the central-peripheral dichotomy (CPD)
theory proposing that vision in the peripheral and central visual fields
are specialized for looking (deciding where to shift the gaze) and seeing,
respectively, and that the feedback query to aid seeing is mainly
directed to the central visual field.  This V1SH-Bottleneck-CPD framework 
predicts that the peripheral visual field, lacking feedback queries, 
is more vulnerable to illusions, and that such illusions become visible 
in the central visual field when the feedback query is compromised.
We present theoretical predictions, experimental confirmations,
a Feedforward-Feedback-Verify-and-reWeight (FFVW) algorithm for seeing 
through the bottleneck, and indicate how the framework
explains visual crowding, grouping, understanding, and post-V1 visual cortical areas.
\end{abstract}

% insert suggested keywords - APS authors don't need to do this
%\keywords{}

%\maketitle must follow title, authors, abstract, and keywords
\maketitle

% body of paper here - Use proper section commands
% References should be done using the \cite, \ref, and \label commands
\section{Formulation of vision in light of an attentional bottleneck}

\subsection{Blind to our own blindness}

If one is born blind, one is unlikely to realize one's own blindness 
until informed  about it by, for instance, a parent.
Even after being informed, it would be difficult for this 
blind individual to comprehend their blindness.  
Surprisingly, we are all victims:
in 1950s, it was observed that human observers can recognize only about 40 bits 
of visual information (an amount enough to encode a short 
sentence of text) per second from about one megabyte (enough to encode a book of text) per second
of visual information sent from the retina to central brain\citep{Sziklai1956,Kelly1962}.
Therefore, we are all blind to more than 99\% of our visual inputs.  
This inattentional blindness was comprehended by few, such that its
demonstrations a few decades later were very striking that, 
 for example, we could not see a gorilla walking into a group of basketball 
players when our attention was occupied with how the basketball moved between the players\citep{SimonsChabris1999}. 

\begin{figure}[ttttthhhhhh!]
\includegraphics[width=5.7 in]{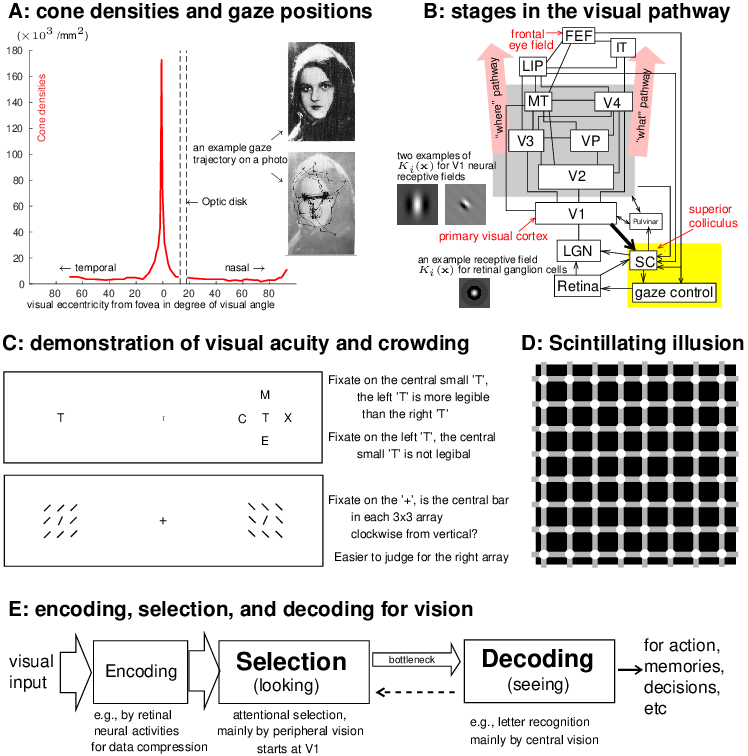}
\caption{\label{fig1} Vision and its formulation. 
A: retinal cone sampling density peaks at the fovea, which is the center of gaze\citep{Osterberg1935}, and
the gaze trajectory of an observer viewing a photo\citep{Yarbus1967}.  
B: along the visual pathway, as visual signals progress from retina, to V1, and further downstream, 
neurons have larger receptive fields. To input image  $S(\bx)$ (a function of location $\bx$),  
the $i^{th}$ retinal or V1 neuron gives responses  
$r_i = f(\int d\bx K_i(\bx -\bx_i) S(\bx) ) +{\textrm {spontaneous response + noise}}$, in which $f(.)$ is
a rectifier or sigmoid-like function. Some examples of $K_i(\bx)$ (for neural receptive fields) are shown.  
C: in the upper panel, the tiny central {\sf `T'} is legible at center of gaze, 
but not when the gaze is directed at the left or right {\sf `T'} (which are 
equally distant from the central {\sf `T'}).  
With the gaze at the central {\sf T}, the right {\sf T} is less legible than the  left {\sf T}
because it is crowded (but not overlapped) by surrounding letters. This suggests 
 that information has been lost 
beyond the retina. In the lower panel, the central bar in the right $3\times 3$ array of bars
is more legible than in the left array when gaze is directed at the `+'.
(Excluding the central bars, the two arrays are left-right symmetric with respect to the `+'.) 
The two central bars are identical to each other, but the right one enjoys a larger
orientation contrast from the surrounding bars, and is thus more salient.
D: illusory gray spots appear in white disks, but only outside the center of gaze.
E:  given the brain's information processing bottleneck, vision is 
formulated as mainly selection (looking) and decoding (seeing), as the 
respective specialisms of peripheral and central vision. 
By contrast, classical receptive fields in retina perform the encoding to compress the data for transmission through the optic nerve 
before the selection\citep{Attneave1954,Barlow1961,ZhaopingBook2014}.
}
\end{figure}

When we fix our gaze at a single word on this page, it is difficult to recognize
a letter several characters away from our gaze.
This is not only because the density of the $10^7$ cones in human retina
decreases from a peak at fovea (Fig. \ref{fig1}A) at the center of our gaze, but also 
because of additional information loss through a processing bottleneck in the brain
that arises beyond the retina.  
This is demonstrated by the upper panel of Fig. \ref{fig1}C,  in which a sufficiently large letter 
in a peripheral visual field location is recognizable when presented in a blank background  
but not when it is surrounded by other letters, even though the surrounding letters
are at least one letter size away and so are unliely to be sampled by the same cones.  
This phenomenon, called visual crowding, manifests our central brain's processing bottleneck. 
This is more apparent in the lower panel of Fig. \ref{fig1}C.
Here, the central bar in the right $3\times 3$ array is more legible than in the
left array, due to its larger orientation contrast from the surrounding bars.  
Critically, neural sensitivity to a bar's orientation is absent in human retina, 
but is present in the primary visual cortex (V1) in the central brain.
Thus, the retina cannot be responsible.
The brain's processing bottleneck is due to an expensive energy 
cost for neural computation\citep{AttwellLaughlin2001}, which means 
that 20\% of the metabolic energy 
in humans at rest is consumed by the brain.

Every second, a human makes about three saccades, which are ballistic gaze shifts 
lasting about $\sim 30$ millisecond (ms).  
However, we have a limited awareness of our saccades, 
such that a typical human reports, when asked, making no more 
than 10 or 20 saccades every minute. 
Meanwhile, everywhere we direct our gaze, we see the letter or object at the center of our gaze 
clearly.  Since most relevant visual information is only one saccade
away from clarity, we have the subjective impression that the 
whole page of  text or the whole scene is clearly seen,
since we fail to realize that the details at a peripheral location 
have been (or will be)  briefly in our center of gaze during a glance. 
We are thus blind to our actual blindness beyond the center of our gaze. 
This is like the impression that the light inside a refrigerator is always 
on since it turns on each time we open the refrigerator's door\citep{Thomas1999,ZhaopingPeripheral2024}.

Not realizing or comprehending our blindness makes it difficult 
to  formulate the problem of vision.  
Consequently, previous theoretical frameworks for vision largely 
ignored this bottleneck. In neurophysiological studies, the focus has been on discovering 
how visual signals are transformed from the retinal photoreceptors to the retinal 
ganglion cells, to neurons in the primary visual cortex (V1) and 
other brain areas downstream from V1 along the visual pathway (Fig. \ref{fig1}B), 
without questioning or examining where along this pathway visual input information 
begins to be deleted.  Psychologists have asked whether the information deletion 
occurs before or after object identification\citep{Broadbent1958, Treisman1985}, 
but without making a link to the physiological substrates.  Progress is difficult 
without a precise theoretical formulation for making non-trivial falsifiable predictions.

\subsection{An information bottleneck starts from the primary visual cortex (V1)}

Researchers swiftly discovered how a retinal or V1 neural responses
$r_i(t)$ (as a function of time $t$  from neuron $i$) depend on retinal input signals
$S(\bx ,t')$ (as a function of visual field location $\bx$ and time $t'$)
\citep{Kuffler1953, HubelWiesel1962, ZhaopingBook2014}.
Non-trivial $r_i(t)$ emerges only for $S(\bx, t')$
within a small visual spatial range $|\bx-\bx_i|$ 
centered at $\bx=\bx_i$ which is known as the neuron's receptive 
field (RF). For retinal and V1 neurons, a RF is typically no larger than the size of a 
small coin at an arm's length.  Neighboring neurons have neighboring and
often overlapping receptive fields, tiling the entire visual field, with 
larger receptive fields in the more peripheral parts of the visual field. 
A retinal or V1 neuron is called a feature detector for 
the optimal feature $S(\bx,t')$ that gives the strongest response $r_i(t)$.
For  example, when time $t$ is omitted for simplicity, 
a simple model for many retinal and V1 neurons is
\begin{eqnarray}
r_i &=& f (L_i)+{\textrm {spontaneous responses + noise, in which}}, \label{eq:LinearRectifier} \\
       && L_i =\int d\bx  K_i(\bx-\bx_i) S(\bx), \nonumber \\
         &&\mbox{and $f(.)$ resembles a rectifier or a sigmoid function}, \nonumber
\end{eqnarray}
so that the optimal feature can be described by $K_i(\bx )$.
This $K_i(\bx )$ for typical retinal neurons resembles a small black/white central dot against a white/black surrounding background 
(see Fig. \ref{fig1}B), and, for  a linear V1 neuron,  it
resembles a small bar or luminance edge (Fig. \ref{fig1}B).  
Noting that a small bar could be made from two neighboring dots, 
and that receptive fields of V1 neurons are somewhat larger than those of retinal ganglion cells, 
one might expect that progressing from retina, to V1, to further downstream areas  (Fig. \ref{fig1}B),
the optimal features become progressively more complex 
(perhaps resembling, e.g., a triangle, square, a face, or a tree), 
while the receptive fields become larger.
However, while the receptive fields do become larger,  
finding the best features to excite the downstream neurons has proved much more difficult, 
despite great advances in experimental methods in recent decades. 
A new framework emerged recently to propose 
that the bottleneck starts from V1's output to downstream visual stages\citep{ZhaopingNewFramework2019}.  
%This bottleneck is consistent with a weaker dependence of downstream 
%neural activities $r_i(t)$ on visual inputs $S( \bx,t'<t)$.

The proposal that the bottleneck starts from V1 is largely motivated
by converging evidence supporting the V1 Saliency Hypothesis (V1SH) proposed two decades ago\citep{LiPNAS1999, LiTICS2002}.
V1SH states that V1 creates (from retinal inputs) 
a saliency map of the visual field to guide gaze shifts before object recognition, so that the highest V1 neural 
response $r(\bx )$ 
to a location $\bx$ (among responses $r_i$ of all the V1 neurons $i$ 
whose receptive fields cover $\bx$)  
represents the value of saliency, defined as the strength of a location to attract gaze 
by visual inputs before object recognition.  It is counter-intuitive that gaze 
can be guided so effectively to the most relevant visual objects for an animal's 
survival before object recognition. The receptive fields of V1 neurons are too small 
to cover an image area sufficient for discerning a face or a predator 
for example.  Meanwhile, logically, the bottleneck precludes 
recognizing all objects in the scene before deciding on the most important
object to which to direct gaze. The brain's solution is for the 
saliency map in V1 to emerge as a global (large scale) effect from short-range interactions 
between local (small scale) receptive fields through neural connections between neighboring 
V1 neurons\citep{Li1997TANC, ZhaopingBook2014} (think about macroscopic 
salt crystals formed by microscopic interactions between microscopic sodium and chloride ions).

Recognizing V1's functional role to guide gaze exogenously 
is essential for understanding downstream visual areas\citep{ZhaopingBrain2016} and
was central to the formulation of the new framework for vision.  
Since shifting gaze should be for the purpose of selecting visual information
into the bottleneck, and since V1 plays an important role to guide gaze (detailed later), 
the bottleneck should start immediately at the output of V1 to the next stage along the visual pathway. 
The bottleneck may be gradual anatomically, and we should find out whether and how 
the massive information loss occurs progressively so that, cumulatively, more information is lost further downstream.

\subsection{Vision as mainly looking and seeing, specialized by peripheral and central vision}

Acknowledging the bottleneck, the new framework formulates the problem of
vision as mainly looking and seeing, so that looking selects a tiny fraction of
visual input information for admission through the bottleneck, and seeing
recognizes the visual objects contained within the selected information. 
Given the superiority of central vision (vision in the central visual field) in seeing, 
this framework naturally includes the central-peripheral dichotomy (CPD) 
theory\citep{ZhaopingFFVW2017,ZhaopingNewFramework2019}. 
The CPD theory proposes that peripheral vision (vision in the peripheral 
visual field) is specialized for looking, particularly for deciding to where in the 
peripheral visual field to make the next gaze shift, and that central vision 
is specialized for seeing.  Henceforth, we call this combination of V1SH, Bottleneck, 
and CPD the VBC framework.

In Fig. \ref{fig1}C, the crowded {\sf `T'} on the right becomes legible when we direct our gaze to it.
This does not  indicate that the processing bottleneck is absent in central vision. 
The CPD theory proposes that the better legibility is not only due to a higher cone density 
in the fovea, but also to a feedback mechanism that enables central vision 
to retrieve additional information through the bottleneck to aid ongoing seeing.
Even without the bottleneck restricting the amount of feedforward information along the visual pathway, 
recognizing the properties of three-dimensional (3D) objects from a two-dimensional (2D) retina image
is an ill-posed problem (since multiple 3D scenes could lead to the same 2D image). 
With the bottleneck, seeing is even more difficult so that perception (the outcome of seeing) can be ambiguous 
(e.g., is this fruit an apple or orange?) or erroneous (in illusions).  
Accordingly, the CPD theory further hypothesizes that additional information is queried from information-richer 
upstream stages (e.g., V1) of the visual pathway by feedback from stages downstream of the bottleneck.  
The queried information (e.g., the color of the fruit) is specific for resolving the ongoing 
perceptual ambiguity (e.g., between an apple and orange) and for vetoing the ongoing illusions.  
Since the feedback query comes after the initial feedforward information flow, it takes 
additional time, such as during the scrutinization of the object.
Since the feedback query is intended to aid seeing, the CPD theory logically 
hypothesizes that it should be mainly directed to the central visual field to save brain resources.  
Lacking this feedback query, peripheral vision is predicted to be more vulnerable
to visual illusions, one such example is in Fig. \ref{fig1}D.

Given the hypothesis that the bottleneck starts from V1's output to 
downstream areas, and given the existing knowledge on the neural receptive 
fields of V1 neurons, precise predictions  can be made for examples of ambiguous perceptions and
illusions in specifically designed conditions. This makes the framework falsifiable. 
Accordingly, as will be detailed later, non-trivial illusions have been predicted to 
be only visible in peripheral vision, which lacks the feedback query to veto the illusions.  
These illusions are perhaps the first non-trivial illusions that have been 
predicted from theories of vision, since, traditionally, visual illusions 
are discovered by accident or by extrapolating from previously known illusions 
and other knowledge.  Furthermore,  such illusions are predicted to be more visible in central vision 
when the feedback query is impaired. We describe confirmations of these predictions below.
Some previously known phenomena and observations, such as visual crowding (Fig. \ref{fig1}C) 
and progressively shrinking coverage of the peripheral visual 
field by visual stages downstream from V1 particularly on the 
ventral visual pathway\citep{ZhaopingBook2014}, can also be understood in VBC terms.

The separation between central and peripheral fields is relative, as they 
really constituting a continuum. Recognizing an object in the peripheral field is possible 
but proceeds less well than in the fovea (Fig. \ref{fig1}C).  
The feedback query for a specific piece of information to aid ongoing seeing 
is a form of looking, when looking is defined as selecting a fraction of 
visual input information into the bottleneck. 
During scrutinization of an object in the center of gaze, 
looking is performed by central vision with minimal or tiny overt gaze shifts 
(more about this later).
After all, the query is to aid seeing, and  the object is already in the center of gaze.
A feedback query can also trigger an overt gaze shift, e.g.,  
from an eye of a face image to the mouth region to perceive 
the facial emotion better (Fig. \ref{fig1}A). 
This gaze shift is partly triggered by the view of the eye in central
vision, and is guided by the knowledge of the general configuration 
of faces and by the relatively peripheral view of the mouth in the specific image.

Before selection in V1, the classical receptive fields of
the retinal ganglion cells perform visual signal encoding (Fig. \ref{fig1}E). 
This encoding compresses (rather than deleting) data (from $10^9$ bits/second available
in the photoreceptors to $10^7$ bits/second across retinal ganglion cells)
to fit visual information for the central brain into the limited channel capacity of the 
optic nerve\citep{ZhaopingBook2014}. 

It is difficult to separate data encoding from data selection strictly, 
since, for example the lower density 
of photoreceptors away from the fovea is already a form of data deletion by sparser sampling.
Meanwhile,  it is also difficult to separate seeing/decoding from mental functions such as reasoning and imagination.
At least in the English language, ``I see" and ``I understand" have similar meanings.  
As will be detailed later, the feedback query to aid seeing involves synthesizing
potential visual inputs based on brain's internal knowledge of the visual world, and this
ability to synthesize is a hallmark of understanding.

The rest of the paper delves deeper into looking and seeing, with derivations of 
the theoretical predictions and their experimental tests, 
and the algorithms realized as neural computation.

%Accordingly, the rest of the paper describes that  peripheral vision is suitably powerful
%for looking and naturally limited for seeing. We will show examples that looking by peripheral vision can occur
%before seeing by central vision\citep{ZhaopingGuyader2007}, and
%can be guided by visual signals that even central vision 
%cannot see\citep{ZhaopingGaze2012,Zhaoping2008OcularSingleton}.
%Then, for seeing, we derive predictions, and their experimental confirmations, 
%of illusions what are visible only in peripheral vision, since central vision has the feedback query to help veto the illusions.
%We show experimental test of thes eWe 
%In contrast, peripheral seeing is vulnerable to visual crowding and misleading 
%visual inputs.  This is largely due to its deficiency in a brain resource
%devoted to central vision: the top-down feedback that aids seeing
%in light of the bottleneck\citep{ZhaopingFFVW2017,ZhaopingNewFramework2019}.

\section{Looking before or without seeing}

Before we focus (in the next section) on seeing through the bottleneck starting 
from V1's output to downstream areas, this section briefly describes that the 
following pre-requirements are satisfied.
Namely, looking can indeed occur before or without seeing, peripheral vision is 
suitably powerful to achieve this, and that V1 is the neural substrates. 
 
\vskip 0.2 in
\begin{figure}[ttttthhhhhh!]
\vfill
\centering
\includegraphics[width=7.2 in]{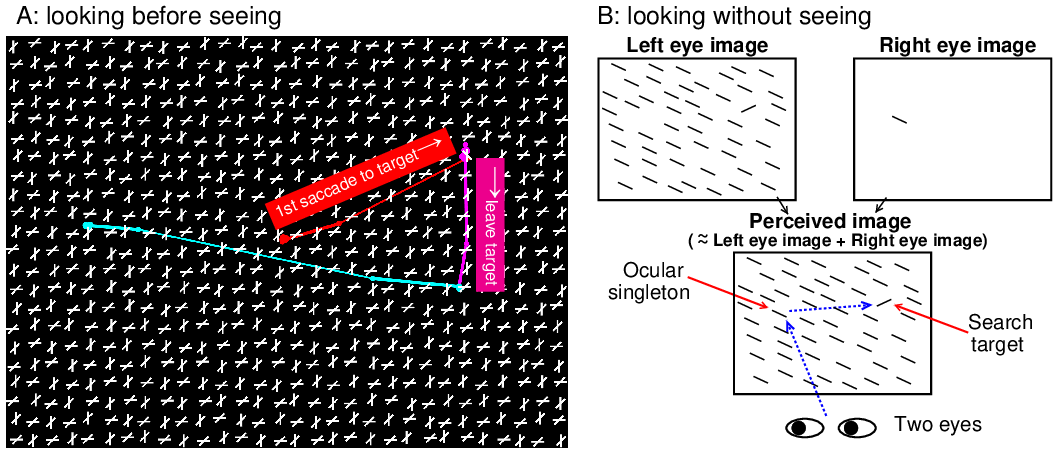}
\caption{\label{fig2} 
Demonstration that looking and seeing are separate processes, and
that looking can occur before or without seeing.
A: on top of a black and white image is superposed  a gaze trajectory 
(together with explanations) in red, magenta, and cyan, from the start, 
to the later moments, of an observer searching in the image 
for a uniquely oriented bar. By design,  the gaze started at the center of the image 
when it  appeared. The first saccade (in red) led the
gaze to the target (here, a bar uniquely tilted counterclockwise from vertical). 
Then gaze stayed around the target for about 0.5 seconds before 
saccading away (in magenta and then cyan) to search elsewhere. 
Visual crowding makes the target bar's unique orientation 
illegible before the first saccade (looking). 
After the first saccade in this example, central vision sees 
the {\sf X} shape made of the target bar and
an intersecting vertical bar. This {\sf X} is a rotated version of all the 
other {\sf X}'s in the image. Rotational invariance in object
recognition confused the observer, leading to the decision to veto the
target and continue searching elsewhere, before returning to the target 
(the returning gaze trajectory is not shown here, for clarity).  
Original data from\citep{ZhaopingGuyader2007}. 
B: looking can occur without seeing, and to a greater degree in more peripheral vision. Observers search
as quickly as possible for an uniquely oriented bar (which differs from 
non-target bars by $50^{\circ}$ in orientation). All bars except one 
are shown only to the left eye, and one non-target bar, the ocular singleton,
is shown to the right eye.  Eye-of-origin of visual inputs is task-irrelevant and
is invisible perceptually; but it is visible to V1, the
only visual cortical area with a substantial number of monocular neurons. The orientation
target and the ocular singleton have the same eccentricity relative
to the center of the perceived image. When this eccentricity is 
$12^{\circ}$ or $7.3^{\circ}$, respectively, 75\% or 50\% of the first
saccades during the search were directed to the
ocular singleton (typically within 300 milliseconds of the appearance of the visual inputs).
Figure adapted from \citep{ZhaopingPeripheral2024}.
}
\end{figure}

In natural behavior, it is not typical to refrain from looking towards an object of interest, 
e.g., the crowded {\sf `T'}  in Fig. \ref{fig1}C, when one tries to see and scrutinize it. 
Logically, looking, especially the first look before any pre-knowledge or 
spatiotemporal context about
the visual scene, must often occur before seeing, since the processing bottleneck 
precludes seeing all sensory inputs in real time before  selection.
However, peripheral vision is suitably powerful for this feat, 
demonstrated in the two examples in Fig. \ref{fig2}.  In Fig. \ref{fig2}A, an 
observer is searching for a uniquely oriented bar as the target, 
but this bar's orientation is illegible 
when his gaze starts the search at the image's center. Nevertheless, the 
first saccade brings the gaze to the target. This saccade is not accidental --
untrained observers bring their gaze to the target within the first second of the search 
in $\sim$ 50\% of the trials in such a search image containing more than 1200 bars. 
In  Fig. \ref{fig2}A, this looking by the first saccade apparently occurred before the 
observer saw the target, since actually seeing the target at the center of gaze 
made the observer erroneously reject it as the target of his search, so that the gaze 
abandoned the target to continue searching elsewhere.
This erroneous rejection arose because the {\sf X} shape made from the target bar 
and an intersecting vertical bar looks identical to other {\sf X}'s in the 
image up to rotation.  Rotational invariance in shape recognition then
confused the observer so that he did not consider this  {\sf X} as special.  
Had the observer recognized the target's  {\sf X}  before his saccade to the target, 
the saccadic plan would be likely cancelled by the confusion. 
If the target bar is tilted only 
$20^{\circ}$ rather than $45^{\circ}$ 
counterclockwise from vertical
to make its associated {\sf X} shape distinctly thinner than the other {\sf X}'s, 
no confusion occurs (as verified in \citep{ZhaopingGuyader2007}).  
Such a search image in  Fig. \ref{fig2}A was specially designed to pit looking 
against seeing, in order to reveal the typically obscured separation between 
looking and seeing in natural vision.

According to V1SH, looking before seeing of this sort is guided by 
intra-V1 neural processing mechanisms\citep{ZhaopingBook2014}. 
V1 does not have neurons whose receptive field $K_i(\bx )$ is 
shaped like the {\sf X}. However, a V1 neuron is activated by a bar in 
its receptive field, when the bar's orientation matches that in 
its $K_i(\bx)$ (see examples in Fig. \ref{fig1}B).  
Let the $i^{th}$ V1 neuron prefer a bar at location $\bx_i$ which is  
tilted at orientation $\theta_i$.  
A critical physiological facet of V1 is 
iso-orientation suppression\citep{AllmanEtAl1985, KnierimVanEssen1992,LinLi1994}, 
by which the activity of this neuron is suppressed 
by each active $j^{th}$ neuron  when $|\bx_i-\bx_j|$ is small and when $\theta_i \approx \theta_j$. 
Iso-orientation suppression reduces V1 neural responses to the horizontal, vertical, and non-target oblique bars 
in Fig. \ref{fig2}A, since each such bars has some neighboring bars tilted in the same 
orientation. The uniquely oriented target bar
escapes iso-orientation suppression, thus evoking the highest response in this image.
This system of recurrently interacting neurons under external visual inputs
is more complex than can be modeled by, or understood as, an 
Ising model\footnote{This is because non-symmetric interactions between excitatory neurons
and inhibitory neurons are mathematically needed in a dynamic system 
to prevent spontaneous breaking of translation symmetry in neural responses 
when visual inputs are translation invariant. This is in order to make the V1 circuit well behaved\citep{LiDayan1999, Li2001, ZhaopingBook2014}.}.
However, numerical and analytical studies confirm that V1's recurrent neural 
circuit can be understood as amplifying V1 neural responses to visual
locations where visual inputs deviate from translation invariance in 
the statistics of visual inputs\citep{ZhaopingBook2014}. 
The {\it local} recurrent interactions between neurons with 
small receptive fields enable the collective behavior of
deciding whether to highlight any visual location (by V1's neural responses) 
depending on the {\it global} context in a visual scene\citep{Li1997TANC,Li1998NC, LiNetwork1999}.

Through V1's known monosynaptic projections to 
the superior colliculus (SC in Fig. \ref{fig1}B), this area's responses are read out as a map.
V1SH states\citep{LiPNAS1999,LiTICS2002}
\begin{eqnarray}
	\mbox{let } r(\bx) &=& \mbox{the highest response
among responses of all V1 neurons whose receptive fields cover $\bx$}, \nonumber \\
	            b(\bx) &=& \mbox{the bid for gaze shifts to location $\bx$}, \nonumber \\
	\mbox{then, } & &\mbox {the psychological quantity $b(\bx)$} = \mbox{$r(\bx)$, a neurophysiological quantity.} \label{eq:V1SH}
\end{eqnarray}
SC mechanisms identify the highest bid $b(\bx)$ across locations $\bx$ to 
direct a gaze shift toward the $\hat \bx$ where $b(\bx)$ is maximum.
According to V1SH, this $\hat \bx$ is simply the receptive field location of the most activated V1 
neuron\citep{Zhaoping2016Evolution,Knudsen2018}.
No seeing or recognition of the  {\sf X} shape 
is needed for this looking by the first saccade in Fig. \ref{fig2}A, 
and indeed no V1 neurons are tuned to {\sf X}. 

In V1, iso-orientation suppression is just one of many forms of iso-feature suppressions; 
these include iso-color and iso-motion-direction 
suppression\citep{AllmanEtAl1985,WachtlerEtAl2003,JonesEtAl2001,ZhaopingBook2014}. 
These are analogous to iso-orientation suppression, except that
the preferred feature value $\theta_i$ is color or motion-direction, rather than (or in addition to) orientation. 
This explains the observations that a unique red apple among green apples is salient 
and so attracts our gaze (attention), as does a bird uniquely flying east among 
a flock of birds flying west. We therefore notice (looking followed by seeing) these 
salient objects even though we are blind to the presence of most other objects in the scene. 

When a uniquely  colored or uniquely moving item attracts gaze towards to it,
the subjective impression is typically that this looking is due to seeing this 
item's perceptual distinction before the gaze shift. 
Separating looking from seeing requires characteristics that are 
either confounded to seeing (as in the rotational invariance of the {\sf X}) or
invisible to seeing. A strong argument for the role of V1 in looking
comes from the fortuitous fact that it 
represents just such a feature. That is  many V1 neurons also prefer the eye of origin 
of visual inputs, so that some prefer inputs from the right eye while some other neurons 
prefer inputs from the left eye.
Iso-eye-of-origin suppression also occurs in V1\citep{DeAngelisEtAl1994}, hence, V1SH predicts that 
an object uniquely shown to one eye among other objects shown to the other eye should be
salient to attract gaze. 
(Stereo goggles for watching 3D movies could be used to show different images to different eyes.)
This is a very surprising prediction, since 
humans are generally unable to perceive whether an object is shown to the left or right eye\citep{OnoBarbeito1985}. 
That is, unlike color and motion direction, eye of origin is not perceptually 
discriminable (unless the two eyes differ substantially by eyesight, lens distortion, etc), 
since all neurons in visual stages downstream 
from V1 are binocular (meaning that inputs from both eyes converge onto each neuron, making the activation of the neuron unable to signal whether the activation is due to input from the left or right eye).
Indeed, it has been known that human observers could not find an item that differs from background
items only in the eye of origin of  inputs\citep{WolfeFranzel1988}. 

Fig. \ref{fig2}B illustrates a confirmation of the resulting 
prediction that an ocular singleton should attract gaze even when it is not perceptually distinctive or discriminable.
In searching for an uniquely oriented bar, observers' initial saccades are
typically toward the ocular singleton, which is a non-target 
in the same scene\citep{ZhaopingGaze2012}. 
By V1SH, both the target and the non-target ocular singleton are salient, by escaping
iso-orientation suppression and iso-eye-of-origin suppression, respectively.
Apparently, the ocular singleton is even more salient, as if it has a unique 
color that is visible only to V1 but not to seeing. 
Control experiments confirmed that observers were unable to tell whether this 
task-irrelevant ocular singleton was present or absent (when each bar is made to have a random luminance value, 
so that observers could not use subtle luminance distinctions 
between the eyes as a cue to identify it)\citep{Zhaoping2008OcularSingleton},
and they are typically unaware of the brief gaze distraction by this ocular singleton during
their search for an orientation singleton\citep{ZhaopingGaze2012,ZhaopingPeripheral2024}.

What may be the ecological utility of making an ocular singleton salient 
to attract looking, while eye-of-origin is not discriminable 
by the subsequent seeing?
By definition, at the location of the ocular singleton there is a strong ocular contrast 
between left-eye and right-eye inputs. In 3D scenes, ocular contrast 
is typically high at borders between a near object and the background 
behind it, since some background content is visible to one eye only at 
such border regions. Attracting gaze to the border helps to select the foreground 
object for the subsequent recognition. Apparently, in Fig. \ref{fig2}B, after 
the ocular singleton bar attracted gaze, although its unique eye-of-origin is not recognized,  
its orientation is recognized to reject it as the search target, thus the gaze shifted 
elsewhere to continue the search.  Apparently, the brain discards 
the eye-of-origin information after V1 after this information has served its function 
for looking via V1.  As we will see later, eye-of-origin information is also essential for determining 
3D depth of objects by binocular correspondence and disparity in stereo 
vision. Discarding this information after V1
apparently does not prevent depth perception by stereo vision.  

Confidence in the veracity of V1SH has been substantially bolstered further by 
two other important pieces of supporting evidence. The first is a direct physiological test of V1SH. 
When a monkey searches for a uniquely oriented bar in a background of uniformly oriented bars, 
saccades to the target with faster onsets are preceeded by  higher initial responses  of V1 neurons to the 
target\citep{YanZhaopingLi2018}. These initial V1 responses
have a latency of 40 to 60 millisecond after the appearance of the bars, and this
latency is too short for the responses to arise
from feedback signals from post-V1 visual areas along the visual pathway.
The second piece of evidence is a behavioral confirmation of a zero-parameter quantitative prediction\citep{ZhaopingZhe2015}. 
The prediction is of the time needed for an observer to find a 
visual feature singleton, and is derived based on equation (\ref{eq:V1SH}), 
pre-existing knowledge about the qualitative properties of neural tuning to visual features 
by V1 neurons but not neurons in post-V1 brain regions, 
and behavioral data on the time needed by this observer to find some 
other types of visual feature singletons.  Confidence in V1SH provides a strong 
foundation for the VBC framework's proposal for understanding seeing after looking.

\section{Seeing through the bottleneck starting from V1}

Seeing is to infer or decode $S$ (here $S$ is for `scene'), i.e., determining the 
object properties (e.g., $S$ is a vector describing color, orientation,
height, number of branches,  leaves and their movements in the wind,  and other information 
about a tree) of a visual scene, from visual input signals 
${\bf r}$. This ${\bf r}$ can be, e.g., an $N$-dimensional vector to represent 
responses from $N$ neurons,  such as cones or V1 neurons responding to the scene.  
The dependence of ${\bf r}$ on $S$  (by image formation and signal transformations 
along the visual pathway) is described by $P({\bf r}|S)$, the conditional probability 
of ${\bf r}$ given $S$.  Through past experience, learning, and evolution, the
brain is assumed to have some knowledge of the 
joint probability $P({\bf r}, S)$ of $\br$ and $S$, and thus also the conditional 
probabilities $P({\bf r}|S)$ and  $P(S|{\bf r})$ (the probability of $S$ given $\br$), 
and the prior probability $P(S)$ of $S$.  
For simplicity of notations, all the probabilities are specified 
by notations for the variables rather than the functional notation $P(.)$.  

Let $\hat S$ denote the decoded value for $S$ in visual perception.
Then, perception as the outcome of seeing should give\citep{ZhaopingBook2014}, 
\begin{equation}
	\text{ideally, perceived $\hat S$}= \mbox{the $S$ to maximize $P(S|{\bf r})$ given encoding responses ${\bf r}$ of all V1 neurons}.  
\end{equation}
When $S$ and ${\bf r}$ are high dimensional vectors, the computational problem 
of computing $P(S|{\bf r}) \propto P({\bf r}|S) P(S)$ 
is subject to the curse of dimensionality\citep{ZhaopingBook2014}. 
For example, if $S$ is an $n$ dimensional vector, and in its $i^{th}$ feature
dimension it could take one of  $m_i$ possible feature values,  then
$S$ could have $\Pi_i m_i$ possible values, and
seeing $S$ requires extracting up to $\sum_{i=1}^n \log_2 m_i$ bits of information 
from $\br$.   If a brain's processing bottleneck allows seeing 
only $C$ bits per second, seeing $S$ could take up to $\sum_{i=1}^n \log_2 m_i/C$ seconds. 
(This is already ignoring the cost of brain's memory space needed to 
store the knowledge $P(\br,S)$ and the cost of retrieving it for this decoding.) 
To reduce the time for seeing, one could reduce the dimension of $S$, 
for example decoding just the orientation, but not the color or 
other properties of an object, and reducing
the resolution of the feature dimension, e.g., only two possible 
feature values of  a particular feature dimension. 
Hence, scrutinizing an object takes time. On the other hand, 
a short glimpse is sufficient to enable a human observer to answer
correctly to a question which has only two possible answers, 
such as, e.g., ``is this bar tilted clockwise or counterclockwise 
from vertical?". A correct answer  ($S=\textrm{``yes"}$ or $S=\textrm{``no"}$) to such an 2-alternative-forced-choice
(2AFC) question contains a maximum of only 1 bit of information.

In practice, for example, when observers need to give an 2AFC
answer to a question ``Is there an animal in this photo?", it takes about $0.5$ second 
for them to report correctly in most trials (although viewing the photos for $20$ ms is 
sufficient)\citep{ThorpeEtAl1996}. With our bottleneck's capacity to process for 
$10^2$ bits of information per second, $0.5$ second (which includes the time to execute the 
reporting action after brain's processing for seeing) could allow more than 1 bit of information.
We assume that this processing complexity increases with the number of
bits $b_{\br}$ needed to represent  ${\bf r}$.
If computing $S$ from $\br$ uses an algorithm like a linear scanning through a look-up table 
of $2^{b_\br}$ possible $\br$ values, the computing time would scale with $2^{b_\br}$; 
a more efficient algorithm could make the computing time scale with $b_\br$ instead. 
Meanwhile, a large $b_\br$ is needed for a ${\bf r}$ for activities of  $10^8$ V1 neurons!

Typically, observers require at least several practice trials 
before doing this  time-constrained task correctly in most of their 
trials. A simple model is that, through instructions and the practice trials, 
the brain learns to narrow down to just $N \ll 10^8$ most relevant V1 
neurons, such that, 
\begin{equation} 
\mbox{practically, brain uses } {\bf r} = (r_1, r_2, ..., r_N) \mbox{~from $N\ll 10^8$ V1 neurons for a non-complex seeing task,} \label{eq:task_set}
\end{equation}
and most likely uses a low resolution on each $r_i$, in order to reduce $b_\br$.
For example,  making $r_i = 0$ or $1$ to signal 
whether neuron $i$ has an activity above a threshold (which could vary with
$i$) would enable $\br$ to convey  up to $b_\br = N$ bits of information. 
Narrowing down to $N\ll 10^8$ neurons and with a low resolution
for each $r_i$ should cut down the complexity for computing $P(S|\br )$.   
Identifying the $N$ neurons on which to focus (and the resolution for each $r_i$) 
for this task requires additional bits of information; this can be done
by additional cognitive processes to set up the so-called task set.
Hence, by setting  the $\br$ in equation (\ref{eq:task_set}), 
the task set prepares for, and controls the execution of, the decoding task. 
Learning and practice trials help to set up this task set 
(such that it costs time and accuracy of task performance when observers switch from performing  
one task to another task\citep{AllportEtAl1994}). 
In any case, the task could be done faster and/or with reduced performance 
if the task set reduces $N$ and and lowers the resolution on $r_i$.  
It is expected that $b_\br$ needed to decode $S$ 
should increase with the number of bits for, and thus the decoding precision for, $S$.
Therefore, the information bottleneck for $S$ in terms of the 
number of bits to represent $S$ should fundamentally reflect the
bottleneck in computational processing for decoding $S$, and
the $b_\br$ for $\br$ needed for the task.

For our current formulation, we simplify this model by 
treating how the task set is set up as a separate question outside our immediate concern.
Then, with our knowledge about how V1 neurons respond to visual inputs, 
this simple model in equation (\ref{eq:task_set}) 
enables the VBC framework to make precise falsifiable predictions.

\subsection{Visual illusions in peripheral vision predicted by the central-peripheral 
dichotomy theory under the bottleneck}

\begin{figure}[ttttthhhhhh!]
\vfill
\centering
\includegraphics[width=6.8 in]{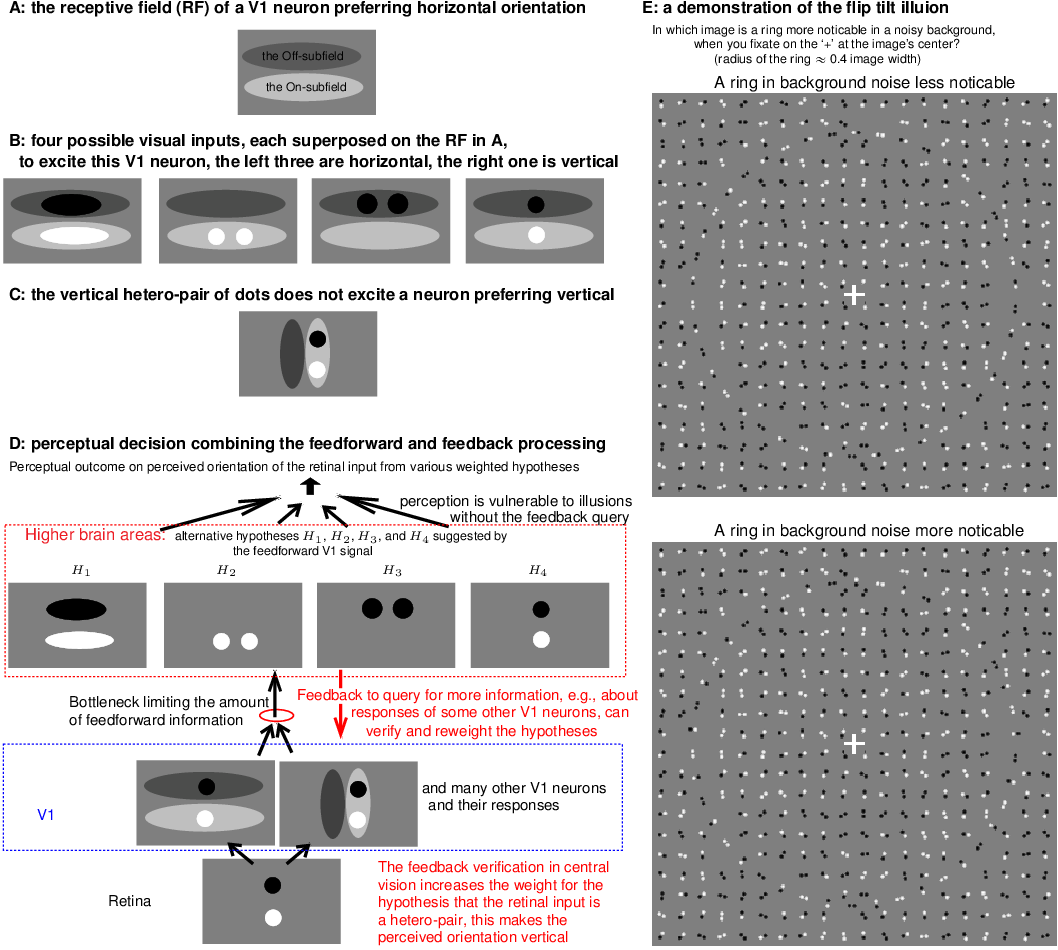}
\caption{\label{fig3}
Seeing  with or without a feedback query to veto an illusion arising from 
impoverished information that is transmitted through the bottleneck. The figure 
illustrates the flip tilt illusion.
A: schematic of a receptive field (RF) of a V1 neuron tuned to a horizontal orientation.
B: four possible inputs $S(x,y)$ to the RF that would excite this neuron: a 
horizontal gabor pattern, a horizontal homo-pair of white dots, a horizontal homo-pair 
of black dots, and a vertical hetero-pair of dots.  
C: the vertical hetero-pair does not activate another V1 neuron preferring vertical. 
D: with the vertical hetero-pair as the retinal input, higher brain areas
	receiving inputs from just the two orientation-tuned V1 neurons would 
suffer the flip tilt illusion $\hat S_{ori}$ that the visual input was horizontal. A feedback query for additional signals 
from other task-relevant V1 neural responses can veto this illusion.
E: a demonstration of this illusion. By a casual look, or with gaze fixated on 
the central `+' in each image, a ring of homo- and hetero-pairs is more easily seen 
in the lower image.  The two images differ only in the orientations of the hetero-pairs 
on the ring. These hetero-pairs are tangential and orthogonal to the ring in the upper and lower images, respectively,
but generate the illusion in peripheral vision of being orthogonal and tangential to the ring. 
This makes the lower ring more noticeable.  When gaze is directed at the ring segments to view them in
central vision, this illusion disappears as the dot pairs 
appear to form smoother and more noticeable segments of the ring in the upper image.
}
\end{figure}

Two particular visual illusions predicted by the VBC framework are the flip tilt illusion 
and the reversed depth illusion. These concern the perception of orientation and depth, respectively, 
which are properties that are signalled by V1 neurons activated by their preferred orientation or 3D depth of visual inputs.
Let image locations $\bx = (x, y)$ have horizontal and vertical coordinates $x$ and $y$. 
If the $i^{th}$ V1 neuron (a simple cell) has a receptive field centered 
at $(x_i, y_i)$ and prefers a horizontally oriented bar or luminance edge, 
its response can follow that in equation (\ref{eq:LinearRectifier}) with a gabor function 
kernel
\begin{equation}
K_i(x, y) 	
\propto \exp \left (-{x^2 \over {2\sigma_x^2}} - {y^2 \over {2\sigma_y^2}} \right ) \cos( k y + \phi ),
\label{eq:Kernel}
\end{equation}
to model the receptive field, with a size and shape 
parameterized by $(\sigma_x, \sigma_y, k, \phi )$. 
Fig. \ref{fig3}A shows an example of such a receptive field (RF) with $\phi = 90^{\circ}$.
The RF regions with $K(x, y)>0$ or  $K(x, y)<0$ are called the on- or 
off-subfields of the RF.  Both subfields  are oriented horizontally, 
so a white horizontal bar falling on the on-subfield can activate this
neuron, as can a black horizontal bar falling on the off-subfield. 
Replacing $\cos( k y + \phi )$ in $K(x,y)$ by $\cos( k x + \phi )$ models another 
V1 neuron tuned to vertical orientation. Many V1 neurons are like a complex cell, whose
response follows equation (\ref{eq:LinearRectifier}) 
after replacing $f(L_i)$ by $f(L^2_{i, 1} + L^2_{i, 2})$, 
with $L_{i, 1}$ and $L_{i, 2}$ involve two different kernels  $K_{i, 1}(x, y)$ and $K_{i,2}(x, y)$ that
follow (e.g.,) equation (\ref{eq:Kernel}) and differ only by having their $\phi$'s $90^{\circ}$ 
apart from each other\citep{ZhaopingBook2014}.
As a result, complex cells are tuned to orientation like simple cells but
are more invariant to the exact location of the bar or luminance edge within their receptive field.

In a 2AFC task to report whether something is oriented vertically or 
horizontally, one of the simplest task sets is to examine responses 
${\bf r} = (r_1, r_2)$ from two V1 neurons whose RFs cover the relevant visual 
location. One neuron, with response $r_1$, prefers, or is activated by, 
horizontal inputs; and the other $r_2$ by vertical inputs. 
Let an object at a particular orientation $S_{ori}$, which can only
be horizontal or vertical, give a visual input pattern $S(x,y)$ (which depends on this object's shape).
Its evoked response ${\bf r}$ has a probability $P({\bf r}|S(x, y))$ 
according to the RF properties or the orientation preferences of V1 neurons. 
Focusing on the task relevant orientation 
feature $S_{ori}$, we have  conditional probability of ${\bf r}$ given $S_{ori}$ as
\begin{equation}
P({\bf r}|S_{ori})) = \sum_{S(x, y)} P({\bf r}|S(x, y)) P(S(x, y) |S_{ori}).
\end{equation}
The optimal perceptual decision
should be  the perceived orientation $\hat S_{ori} =h$ (horizontal) or
 $\hat S_{ori} =v$ (vertical) according to\citep{ZhaopingBook2014}
\begin{equation}
\hat S_{ori} =
\left \{
\begin{array}{ll}
	      h, & \text{if ~~} P(S_{ori}=h|{\bf r})> P(S_{ori}=v|{\bf r}), \\
	      v,& \text{otherwise}.
\end{array}  \label{eq:MAX_ori}
\right.
\end{equation}
In many simple situations such as 
when all of the following four conditions hold: 
$P({\bf r} =(\alpha , \beta ) |S_{ori} = h) = P({\bf r} =(\beta , \alpha) |S_{ori} = v)$
for any $(\alpha, \beta)$, 
$P({\bf r}|S(x,y)) =\Pi_i P(r_i|S(x,y))$, $P(r_i|S(x,y))$ is Poisson,  
and $P(h)=P(v)=0.5$,
%when $P(r_1=i,  r_2 |S_{ori} = H) = 1$ and $P(r_1< r_2 |S_{ori} = V) = 1$ both hold, 
%or when $P({\bf r}|S(x,y)) = P(r_1|S(x,y))P(r_2|S(x,y))$ is Poisson for each $P(r_i|S(x,y))$ 
%and $P(H)=P(V)$ both hold, 
equation (\ref{eq:MAX_ori}) can be shown\citep{ZhaopingBook2014} to be equivalent to
\begin{equation}
\hat S_{ori}  = h \text{~or~} v \text{~~if a decision variable~~ }
                     R \equiv r_1 - r_2 >0 \text{~~or otherwise.}  \label{eq:R}
\end{equation}
In neural hardware, this perceptual decision process could be easily 
implemented by a feedforward network having a decision neuron downstream from V1 receiving excitatory
input signals from $r_1$ and inhibitory input signals from $r_2$.  
Hence, for example ${\bf r} = (1,0)$ or $(0,1)$ gives a perception of 
horizontal or vertical orientation, respectively. 

Let $r_1$ arise from a RF  $K_1(x, y)$ depicted by Fig. \ref{fig3}A. 
Fig. \ref{fig3}B shows four possible visual inputs $S(x, y)$ to
give a substantial response $r_1$. The first $S(x, y)$ is a gabor pattern 
resembling $K_1(x, y)$; 
the second or third $S(x,y)$ depicts a horizontal homo-pair of dots, 
both dots are white or both dots are black. 
These first three $S(x,y)$'s have horizontal orientation $S_{ori} = h$.
However, the last $S(x,y)$ depicts a vertical hetero-pair of dots, 
one black and one white.  This vertical hetero-pair, with $S_{ori} = v$, 
nevertheless evokes a non-trivial $r_1$ by having the black and white dots inside the off- and on-subfields. 
However, it does not excite a vertically tuned RF $K_2(x, y)$, giving $r_2 = 0$ (Fig. \ref{fig3}C).
By the decision variable $R$ in equation (\ref{eq:R}), this hetero-pair of dots 
should evoke an illusion of horizontal orientation.  
This flip tilt illusion\citep{ZhaopingFlipTilt2020}
makes the perceived $\hat S_{ori}$ equal to the outcome of flipping the 
actual $S_{ori}$ by $90^{\circ}$.
Readers can experience this illusion in Fig. \ref{fig3}E.

However, through prior knowledge (much presumably gained from previous experience), 
the brain knows the joint probability $P(S(x, y), {\bf r}, S_{ori})$ of 
$S(x, y)$, ${\bf r}$, and $S_{ori}$ (and hence the relevant conditional probabilities).
Hence, from the  ${\bf r}=(r_1, r_2)$ admitted through the bottleneck for seeing, the task set can assess, 
via $P(S(x, y)|{\bf r})$, all the possible $S(x, y)$ that are consistent with  this  ${\bf r}$.
Each possible $S(x, y)$ is a perceptual hypothesis, denoted as $H_j$, for $j = 1, 2, ...$, 
about the sensory input scene $S(x, y)$ (a particular object at a particular 
orientation $S_{ori}$) as a cause of ${\bf r}$.  For example, in the toy model Fig. \ref{fig3}D, there are four hypotheses, 
$H_1$ for the gabor pattern, $H_2$ for the white homo-pair, 
$H_3$ for the black home-pair, and $H_4$ for the hetero-pair, that could 
cause ${\bf r} = (r_1, r_2) = (1, 0)$.  
Since $S_{ori}=h$ for $H_1$, $H_2$, and $H_3$ but
$S_{ori}=v$ for $H_4$, 
then, 
\begin{eqnarray}
P(S_{ori}=h |{\bf r}) &=&  \sum_{j=1}^3 P(H_j|{\bf r}) \propto \sum_{j=1}^3 P({\bf r}|H_j) \cdot P(H_j), \\
P(S_{ori}=v |{\bf r}) &=&   P(H_4|{\bf r}) \propto  P({\bf r}|H_4) \cdot P(H_4).
\end{eqnarray}
Hence, $P(S_{ori}=h|{\bf r}) > P(S_{ori}=v|{\bf r})$ assuming 
$P({\bf r}|H_j)$ is the same for all $j = 1, 2, 3, 4$ and $\sum_{j=1}^3 P(H_j)> P(H_4)$ 
from the brain's knowledge of the statistics of visual scenes. 
By equation (\ref{eq:MAX_ori}), the perceived orientation is $\hat S_{ori}=h$.
In natural scenes, neighboring pixel values are correlated, making it most likely 
that $P(H_i) > P(H_4)$ for each $i< 4$.  This is likely why the flip tilt illusion is quite 
strong in Fig. \ref{fig3}E. 

Meanwhile, the probability that $\hat S_{ori}=h$ is erroneous is $P(S_{ori}=v|{\bf r})$, 
which is non-zero, albeit less than the probability of decoding error if $\hat S_{ori} = v$ instead.
To verify $\hat S_{ori}= h$, the task set could send a feedback query to V1 for 
additional information, e.g., about the responses from some other neurons 
or for a finer resolution version of $r_1$ and $r_2$. 
From $P(r_k| H_j)$, the task set could generate or synthesize likely 
$r_k$ values for each $H_j$ for $k=1, 2, 3, 4 ...$. 
Let 
\begin{eqnarray}
\hat r_k (H_j) &=& \text{the synthesized would-be response of the $k^{th}$ upstream neuron if $H_j$ holds for the scene,}\\
r_k  &=& \text{the actual response from the $k^{th}$ upstream neuron.}
\end{eqnarray}
The  synthesized $\hat r_k (H_j)$ is fed back from downstream to upstream 
areas along the visual pathway to compare with the actual $r_k$. 
A hypothesis $H_j$ is vetoed if $\hat r_k (H_j) \not\approx r_k$.
The task set should choose the set of $k$ that best discriminate between the 
alternative $H_j$'s to resolve the ongoing ambiguity, as the bottleneck 
should preclude querying for responses of too many neurons. 
For example, for the toy model in Fig. \ref{fig3}D, the task set 
could query for ${\bf r'} = (r_3, r_4)$ from two 
additional neurons whose receptive fields cover the same location but whose 
preferred orientations are $45^{\circ}$ clockwise and counterclockwise 
respectively from vertical.
When a $H_j$ is vetoed because  $(\hat r_3 (H_j), \hat r_4 (H_j)) \not \approx (r_3, r_4)$, 
then $P(H_j|(r_1, r_2, r_3, r_4) \approx 0$. In Fig. \ref{fig3}D, such a query should 
give $P(H_4|(r_1, r_2, r_3, r_4) \approx  1$, and consequently,  
the illusion $\hat S_{ori}= h$ is vetoed.

Fig. \ref{fig3}E demonstrates that the flip tilt illusion, manifested
as the higher noticeability of the ring in the lower than upper
images,  is absent in central vision when one directs the gaze to 
the ring segments. This is consistent with the CPD theory that
the feedback query for additional information to aid seeing
is mainly for central vision.

The feedback query is an active process, and it is part of a seeing algorithm called 
Feedforward-Feedback-Verify-and-reWeight 
(FFVW)\citep{ZhaopingFFVW2017, ZhaopingNewFramework2019}.
Using the example in Fig. \ref{fig3}D to illustrate, FFVW has the following
steps.

\begin{itemize}

\item First is the feedforward step. 
Selected upstream visual signals ${\bf r}$, 
e.g.,  ${\bf r} = (r_1, r_2)$ from V1 neurons, 
provide feedforward information to downstream visual stages, suggesting initial hypotheses, e.g.,  
$H_1$, $H_2$, $H_3$, and $H_4$, for visual object properties, 
each $H_j$ has a weight for its likelihood to be correct as
\begin{equation}
w_j  \propto P({\bf r}|S(x, y) \textrm{ for } H_j) \cdot P(H_j), \mbox{~~so that $\sum_j w_j = 1$.} \label{eq:HjWeights} 
\end{equation}
In the toy example, this could give, e.g., $w_j = 0.25$ for each $j=1, 2, 3, 4$ of the four hypotheses.

\item
Second, the feedback step. From each $H_j$, the downstream stages 
use knowledge $P(r_k| S(x, y)  \textrm{ for } H_j)$ to generate or synthesize selected
would-be upstream neural responses ${\hat {\bf r}}'(H_j)$ if $H_j$ holds. 
The actual upstream responses ${\bf r'}$, 
e.g., ${\bf r'} = (r_3, r_4)$ from V1 neurons, should contain information that is not yet in the 
already received responses ${\bf r}$ in the downstream stages.  
The ${\bf r'}$ may be responses from neurons other than those 
that gave $\br$, or may be from the same neurons as the ones that give $\br$ but 
the query seeks for more details, or a higher resolution version, of $\br$.
The task set makes the choice of which neurons' responses to query, so as to
best discriminate between the alternative $H_j$'s.
The downstream stages feedback the would-be 
 ${\hat {\bf r}}'(H_j)$ to compare with the actual ${\bf r'}$ in the upstream stages such as V1. 

\item
Third, the verification step, to verify whether ${\hat {\bf r}}'(H_j) \approx {\bf r'}$.

\item Fourth, the reweighting step. The degree of match between 
the would-be   ${\hat {\bf r}}'(H_j)$ and actual ${\bf r'}$ is used to modify the weights $w_j$ 
so that a good or poor match increases or decreases $w_j$ accordingly.
The updated weights are expected to approach those according to equation \ref{eq:HjWeights} 
after replacing the initial feed forward signals $\br$, e.g., ${\bf r}=(r_1, r_2)$, by 
expanded signals $(\br, \brp )$, e.g., $({\bf r, r'})=(r_1, r_2, r_3, r_4)$. 

\item Together, the previous four steps could be iterated, depending on the accuracy 
and speed needed for the decoding and on the neural resources available. 
In each iteration, the $\br$ in its first step includes the initial 
feedforward signals and the signals already queried by previous iterations if any,
the $\brp$ in its other steps is the queried signals by the
current iteration.
\end{itemize}

This FFVW algorithm paraphrases the long-standing 
analysis-by-synthesis\citep{Neisser1967, CarpenterGrossberg1987ART, KerstenMamassianYuille2004, YuilleKersten2006}, 
designed to analyze sensory inputs for sensory inference (decoding) by synthesizing 
the would-be sensory signals consistent with potential outcomes of the inference.

In peripheral vision, a lack of neural resources for the feedback 
query (according to the CPD theory) nullifies FFVW, so that
it simplifies to just a Feedforward-and-Weight (FfW) algorithm. 
This FfW algorithm under the information bottleneck makes visual perception
ambiguous, as manifested in visual crowding (demonstrated in
Fig. \ref{fig1}C), and makes perception vulnerable to illusions, 
as demonstrated in Fig. \ref{fig1}D and Fig. \ref{fig3}E.

By contrast, in central vision, perceptual ambiguity can normally be resolved and 
illusions vetoed thanks to the  FFVW algorithm.
Since the bottleneck allows past only a tiny fraction of visual 
input information per second, the amount of information that is queried 
has to be sufficiently small. Hence, the queried information has to be selective, 
so as not to waste the bottleneck's capacity on less relevant information, such as 
that about the responses of the V1 neurons whose receptive fields cover 
locations too far from locations of interest.
If visual input information were passively and non-selectively accumulated 
at downstream stages at a rate limited by the transmission 
capacity of the bottleneck, it would perhaps take minutes, if not hours,
 before there is a substantial chance for the task 
relevant information to be available to impact the perceived $\hat S$. 
This even assumes that the retinal input image was
relatively static (and not removed) during the information accumulation.
This is consistent with the observation that visual crowding in the 
peripheral visual field is not substantially reduced by merely viewing the visual inputs 
for much longer than $0.1$ second.

It should be noted that the predicted flip tilt illusion in 
peripheral vision and its invisibility in central vision are 
not dependent on how the task set has been set up, and whether the queried 
responses are $(r_3, r_4)$ or something else.
The predictions can be derived as long as we have:
(1) the  known properties of  receptive field shapes of V1's orientation tuned neurons;
(2) the presence of a bottleneck starting from V1's output to downstream areas 
to limit the initial $\br$ used for decoding $S_{ori}$ (so that the decoding outcome 
could be erroneous);
(3)  a sufficient amount of additional task-relevant information available in 
	the responses of the whole population of V1 neurons to the scene, 
		beyond the information already sent forward in $\br$;
(4) the availability of the feedback query to central but not peripheral vision
to access the additional task-relevant information in V1 to resolve the perceptual ambiguity and veto the illusion.

\begin{figure}[ttttthhhhhh!]
\vfill
\centering
\includegraphics[width=6.0 in]{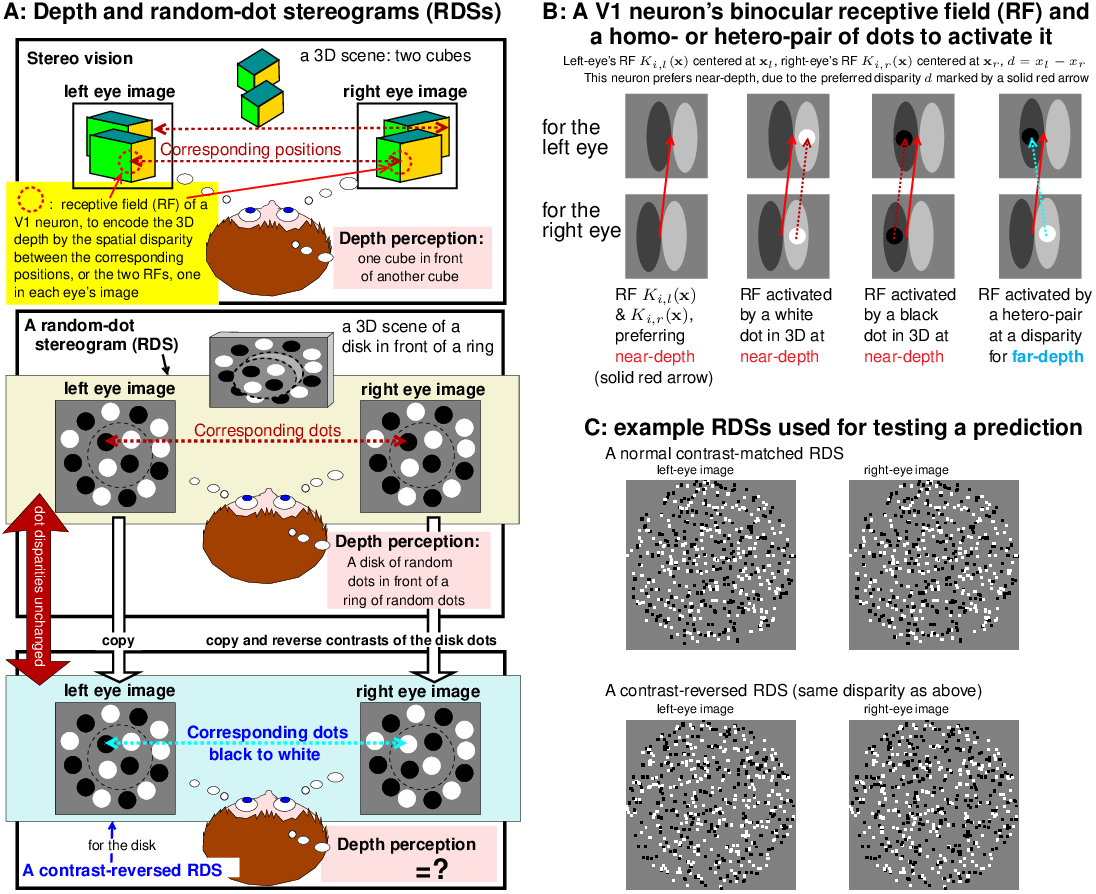}
\caption{\label{fig4}
Depth perception, random-dot stereograms, and the reversed depth illusion through V1 
neural signals.
A: 3D signals are encoded by  V1 neurons with receptive 
fields (RFs) in both eyes. Typically, such a neuron has two similar monocular RFs.
The two RFs prefer spatial locations $(x_l, y_l)$ and $(x_r, y_r)$ in the left-eye 
and right-eye images, making the neuron prefer a 3D depth from its preferred 
binocular (horizontal) disparity $x_l-x_r$. 
If 3D objects are visible only by the random black and white dots on their surfaces, 
their images on the two retinas constitute a random-dot stereogram (RDS).   
The depth of a  surface, e.g., a disk or a ring, is signalled by 
the depth-tuned V1 neurons responding to the random dots.
If the stereogram is modified by flipping the contrast-polarity of 
each dot for one surface (e.g., the disk) in one monocular image (e.g., the right-eye image),
the stereogram is called a contrast-reversed RDS. The affected surface is non-sensical,
since each black dot in one eye corresponds to a white dot in the other eye for this surface.
B: a schematic illustrating that the preferred depth, near or far, 
of a V1 neuron to a normal 3D dot becomes anti-preferred for a 
non-sensical dot made of a dichoptic hetero-pair of dots, 
as one from a contrast-reversed RDS.
For illustration, we use a simplistic model of $i^{th}$ V1 neuron's receptive field,
made of two monocular RFs, $K_{i, l}(\bx )$ and $K_{i, r}(\bx )$ for the left-eye and 
right-eye images. The two monocular RFs have the same shape but are 
spatially displaced from each other by $d = x_l - x_r$.
To normal stereograms, the neuron prefers near or far depth when the 
preferred disparity $d$ satisfies $d>0$ or $d<0$. 
C: two example RDSs depicting the same 3D scene of a disk in front of
a surround ring. The upper RDS is a normal RDS, the lower one
is a contrast-reversed RDS, so that, for the disk, a black dot in one eye
corresponds to a white dot in the other eye. Free fusing should enable you to
see the disk and the ring in the upper RDS, but the depth order is completely 
unclear in the lower RDS unless the RDS is viewed at a more peripheral visual location 
to enable a vague perception of the reversed depth illusion that the disk is behind the ring.
}
\end{figure}

Illusions of motion direction and 3D depth analogous to the flip tilt illusion
are called the reversed phi motion illusion\citep{Anstis1970} and the reversed depth 
illusion\citep{ZhaopingAckermann2018}.  In the same way that a homo-pair of two dots 
at locations $(x_1, y_1)$ and $(x_2, y_2)$ defines an orientation according to
$(x_1-x_2, y_1-y_2)$, a homo-pair of two dots in space-time 
$(x_1, y_1,  t_1)$ and $(x_2, y_2, t_2)$ defines a motion direction from 
$(x_1, y_1)$ to $(x_2, y_2)$ (which can be perceived as apparent motion when 
the spatiotemporal difference between the two dots is sufficiently small). 
Analogously, a homo-pair of two dots, one shown to the 
left eye and another to the right eye, can define the depth of one dot in 3D space 
imaged at retinal image locations $(x_l, y_l)$ and $(x_r, y_r)$ in the left and right eyes.
The depth of this dot relative to the fixation location in 3D is
defined largely by the horizontal disparity $x_l - x_r$ (see Fig. \ref{fig4}A).  
In V1, in addition to neurons tuned to orientation, there are  neurons 
tuned to motion direction by their spatiotemporal RF $K_i(\bx, t)$ 
and neurons tuned to depth (or disparity) by two RFs,
$K_{i, l}(\bx )$ and $K_{i, r}(\bx )$, 
for the two eyes for each depth-tuned neuron $i$\citep{ZhaopingBook2014}.
When the homo-pairs are replaced by hetero-pairs of dots as visual inputs, 
the preferred motion direction or preferred depth of V1 neurons also flips to the opposite 
motion direction\citep{PriebeFerster2005} or the opposite depth (nearer versus farther from the 3D 
fixation location)\citep{CummingParker1997} (see Fig. \ref{fig4}B), analogous to the flipping of 
neural preferred orientation schematized in Fig. \ref{fig3}. 

In human perception, the reversed phi motion illusion has long been observed and 
is indeed stronger in peripheral vision\citep{Anstis1970}. 
It had long been thought that the reversed depth illusion is 
invisible\citep{CummingEtAl1998}, a conclusion reached by  studies that examined only 
central vision since peripheral vision was traditionally viewed as only quantitatively 
different from central vision (mainly by a lower spatial resolution). 
This invisibility of V1 signals was rationalized by 
an assumption that visual awareness is outside V1\citep{CrickKoch1995}.
This changed when the CPD theory predicted, with experimental confirmation, 
that reversed depth is indeed visible in peripheral vision\citep{ZhaopingAckermann2018}.

\subsection{Reversed depth illusion becomes visible in central vision when the feedback query is impaired}

That the flip tilt and reversed depth illusions are visible to peripheral, but not central, 
vision constitutes strong support to the CPD theory. 
However, an alternative explanation for the invisibility of these illusions 
in central vision could be the higher density of retinal cones for central vision,
and correspondingly a larger number of V1 neurons for each unit of solid angle 
of visual space in central than peripheral vision.  
Using the reversed depth illusion, Zhaoping and Ackermann\cite{ZhaopingAckermann2018} indirectly 
argued against this alternative by showing that enlarging the input images in peripheral vision 
does not weaken this illusion.
However, a more direct test against this alternative can be
to test another prediction of the CPD theory: impairing the top-down feedback 
query should make the illusion visible in central vision. 

We can test this prediction using the reversed depth illusion. 
Random-dot stereograms can depict 3D scenes using exclusive stereo cues.
An example is the scene containing a disk in front of or behind a surrounding ring  
schematized in Fig. \ref{fig4}A. 
The shape of the disk and the ring are not discernible in either 
monocular image (see the actual examples in Fig. \ref{fig4}C).
However, the spatial correspondence between the two dots in the two retinas
projected from a single dot in 3D space provides a 3D depth cue 
encoded by depth-tuned V1 neurons. This cue is the spatial 
disparity  $d\equiv x_l-x_r$ between the horizontal image locations $x_l$ and $x_r$ 
of this 3D dot in the left-eye and right-eye monocular images.  
This dichoptic correspondence between 
two dots, one each in the left- and right-eye images,
as signalling depth  is analogous to a homo-pair of dots in 
Fig. \ref{fig3} signalling orientation. 
(Using stereo goggles, observers viewing the upper stereogram in  
Fig. \ref{fig4}C should vividly see a disk in front of a surrounding ring.)  

When the dichoptic correspondence is between a black dot in one eye and a 
white dot in the other eye, it is analogous to a hetero-pair of dots in Fig. \ref{fig3}, 
such that V1 neurons signal the opposite depth to such a non-sensical hetero-pair. 
In other words, to a hetero-pair of disk dots (in a contrast-reversed RDS) 
at disparity $d>0$ in front of a background ring (of zero disparity), 
V1 neurons tuned to the opposite disparity ($d<0$) are activated (Fig. \ref{fig4}B).  
These V1 responses $\br$, from V1's binocular depth-tuned neurons, 
can be fed forward to suggest to downstream visual areas an hypothesis $H_j$ 
that the disk is behind the ring, causing the reversed depth 
illusion seen by peripheral but not central vision\citep{ZhaopingAckermann2018}. 
We may assume for concreteness that two initial hypothesis $H_1$ and $H_2$ are
suggested by the feedforward signals: $H_1$ for the reversed depth with 
a weight $w_1$ and $H_2$ (with a weight $w_2$) for something non-sensical without 
clear depth signals, and that $w_1>w_2$. 
Peripheral vision illusorily sees $H_1$, since $w_1>w_2$, through the FfW algorithm.
In contrast, central vision vetoes $H_1$ via FFVW 
using an additional $\brp$ arising from the feedback query.  
One informative $\brp$ would be the responses of V1 monocular neurons
to the two monocular dots in the hetero-pair. 
The expected ${\hat \br}'(H_1)$ from the reversed depth hypothesis $H_1$, 
which assumes a dichoptic homo-pair of dots, 
would be inconsistent with the actual  $\brp$ to a dichoptic hetero-pair of dots, 
i.e., ${\hat \br}'(H_1)\not\approx \brp$, thus vetoing $H_1$.  
When the feedback query is impaired, central vision is predicted 
to be able to see this illusion from $H_1$ also. 

To test this prediction,  we take advantage of the expectation 
that the feedback query should 
take time. Consider a visual input, such as a RDS that evokes activity in V1
at time $t_1$, along with a feedback query for $\brp$ being
delivered to an upstream stage 
such as V1 along the visual pathway to verify whether ${\hat \br}'(H_j) \approx \brp$.
Imagine that this feedback query arrives $\delta t$ later than $t_1$ at time $t_2 = t_1+\delta t$.
If the original RDS is quickly replaced by another visual input which we call mask 
(this procedure is called backward masking), 
so that  V1 neural responses, including $\brp$, to the RDS 
are replaced  by the responses to the mask before $t_2$, 
the feedback query is prevented.
This requires that the original stereogram be shown for no longer than 
a brief duration $\delta t$. This $\delta t$ could be 
around $30$ to $40$ millisecond according to neurophysiological 
data\citep{ChenYanEtAl2014, ChenWangEtAl2017, YanZhaopingLi2018, KlinkEtAl2017}.

To enable observers to view the RDS for a sufficiently long time, 
we let observers see multiple, successive, frames of RDSs, 
one after another, each for a $10$ ms.
All the RDS frames  are for the same disk and the ring, 
with the same statistical properties such as the (uniform) density of the random dots, the disparity 
of the disk relative to the ring, the sizes and locations of the ring and the disk, 
and whether the RDS is contrast-matched or contrast reversed for the disk 
(the ring is always contrast-matched).  Different frames differ from each other 
only in the exact set of dots randomly distributed on the disk and the ring, 
this set is randomly generated independently between different frames.
Hence, each frame is masked by the next frame, preventing
the feedback query.
When the RDSs were contrast-matched, observers reported the depth order 
between the disk and the ring correctly in almost 100\% of their trials
(the disk was in front of the ring in randomly half of the trials).
Hence preventing the feedback query  does not
impair seeing the depth clearly and correctly when 
there is no illusion to veto.
When the RDSs were contrast-reversed (for the disk), observers
reported the reversed depth order in most trials. 
Hence, they could see the illusion, even though
the RDSs were viewed in central vision, confirming 
the prediction\citep{Zhaoping2021ECVP,ZhaopingVSS2024}.

\subsection{Prioritization of the feedforward visual signals and the feedback queried signals for seeing}

\begin{figure}[ttttthhhhhh!]
\vfill
\centering
\includegraphics[width=6.3 in]{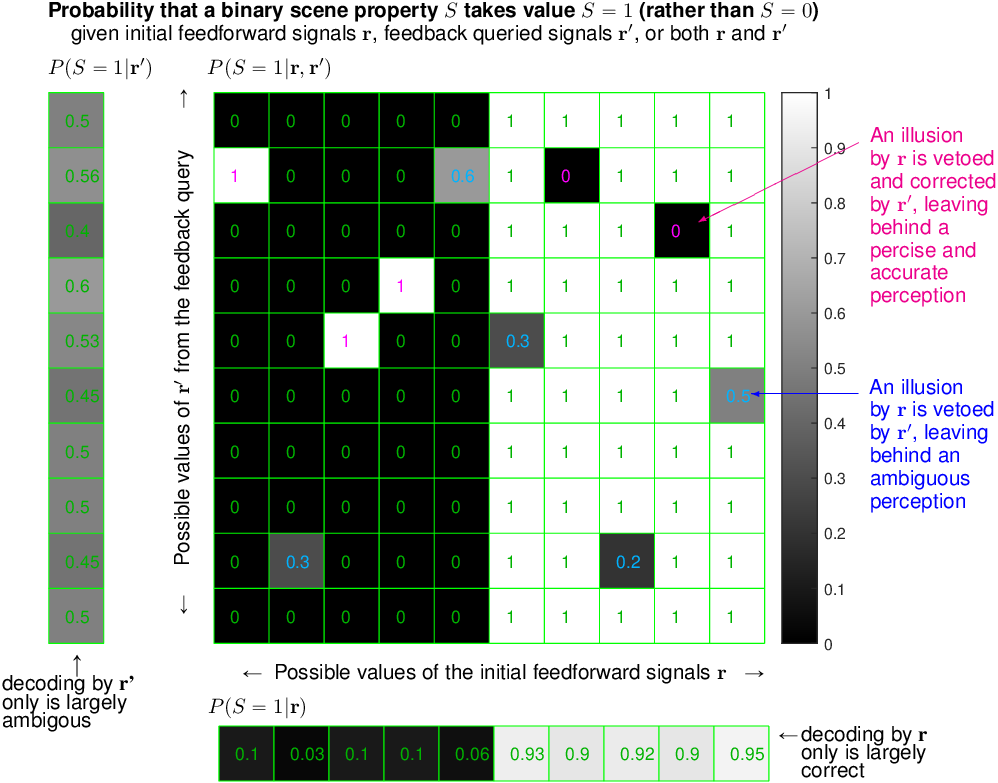}
\caption{\label{fig5}
Decoding  a binary scene property $S$ (e.g., $S=$ whether a disk is in front of a ring) 
from feedforward ${\bf r}$ only,  the feedback queried signals 
${\bf r'}$ only, or both, $({\bf r, r'})$, a toy schematic 
to illustrate perceptual illusions, ambiguities, and the organization of task set for visual recognition.
All discrete (noise-free) $({\bf r, r'})$ values, separated by green borders, 
are equally probable (for simplicity). 
Probabilities $P(S=1|{\bf r})$, $P(S=1|{\bf r'})$, and $P(S=1|({\bf r', r'}))$
are visualized by gray shading and quantified by green, magenta, or blue colored numerical numbers.
Decoding by a $(\br, \brp )$ is most precise when $P(S=1|({\bf r', r'}))=1$ or $0$, 
more ambiguous as $P(S=1|({\bf r', r'})) \rightarrow 0.5$,  analogously for 
decoding by ${\bf r}$ or ${\bf r'}$ only.
A $({\bf r', r'})$ marked by magenta-colored $P(S=1|({\bf r', r'}))=1$ or $0$ 
is when an illusion decoded by ${\bf r}$ only is vetoed and corrected by ${\bf r'}$,
like the veto of the flip tilt illusion by central vision in Fig. \ref{fig3}. 
A $({\bf r', r'})$ marked by blue-colored $0<P(S=1|({\bf r', r'}))<1$ is 
when ${\bf r'}$ vetoes the illusion decoded by $\br$ alone but leaves behind an ambiguous percept,
such as when the reversed depth illusion is invisible in typical central vision in Fig. \ref{fig4}.
Since $\br$ is more informative than $\brp$ about $S$, 
 $\br$ should be preferred over $\brp$ as  the initial feedforward signal for this decoding.
}
\end{figure}

To decode for $S$, by what criteria should the task set use
to choose which feedforward signals $\br$ and feedback queried 
signals $\brp$ to be admitted for decoding? Fig. \ref{fig5} provides a toy example with a binary $S$, 
e.g., about whether a disk is in front of a ring.
Variations of other scene properties $S'$ (e.g., about shape) 
allows each $S$ to evoke multiple possible $({\bf r, r'})$ 
(ignoring random fluctuations in  $({\bf r, r'})$ for simplicity).
The amount of information about $S$ provided by ${\bf r}$ is
quantified as the mutual information (for simplicity, but without lose of generality,
$S$ and $\br$  are variables taking discrete values)
\begin{equation}
	\begin{array}{l}
      I(S; {\bf r}) = H(S) - H(S|{\bf r}), \mbox{~~ in which} \\
	H(S) = -\sum_S P(S)\log_2 P(S),  \mbox{~~and~~}
	H(S|{\bf r}) = -\sum_{S, {\bf r}} P({\bf r}) P(S | {\bf r})\log_2 P(S | {\bf r}). 
	\end{array}
\end{equation}
When  $P(S|{\bf r})=1$ or $0$ for any $S$,    
 using ${\bf r}$ can decode $S$ precisely, with zero uncertainty  $H(S|{\bf r})$, thus  $I(S; {\bf r}) = H(S)$. 

In the example in Fig. \ref{fig5}, each $\br$ gives 
either $P(S=1|{\bf r})\approx 1$ or $\approx 0$ (note that $P(S=0 |\br ) = 1- P(S=1 |\br )$), 
hence, ${\bf r}$ conveys $S$ with only a small uncertainty  
$H(S|{\bf r})>0$. In comparison, the uncertainty  $H(S|{\bf r'})$ by $\brp$
alone (without $\br$) about $S$ is larger since for most ${\bf r'}$, 
$P(S=1|{\bf r'}) \sim  0.5$. Hence, $\br$ is more informative 
than $\brp$ about $S$. If feedback query is infeasible due to a limited processing time 
or feedback resources, it would be an inefficient use of the bottleneck 
to admit $\brp$ rather than $\br$ for decoding $S$ in the initial feedforward 
step (assuming that transmitting $\br$ and $\brp$ consumes the same amount 
of channel capacity). For example, to decode whether the orientation is horizontal 
or vertical in Fig. (\ref{fig3}), having $\br$ as responses from two neurons, one tuned to horizontal
and the other to vertical, is better than from two neurons tuned to $45^o$ from vertical
in opposite directions.
Thus, 
\begin{equation}
\begin{array}{l}
   \text{given a limited capacity (bottleneck) for admitting any signals to decode a scene property $S$,} \\ 
             \mbox{the feedforward $\br$ should be from the most task-relevant neurons, 
		so as to maximize $I(S; \br)$.} \label{eq:Get_br}
\end{array}
\end{equation}

Meanwhile, given $\br$, having 
the feedback queried $\brp$ additionally 
can potentially boost decoding quality.
For example,  ${\bf r'}$ could veto an illusion by $\br$ only 
to give a correct percept (like vetoing the flip tilt illusion by central vision in Fig. \ref{fig3}),
in  Fig. \ref{fig5}, this occurs for $(\br , \brp )$  marked by magenta-colored $P(S=1| {\bf r, r'})$.
Also,  ${\bf r'}$ could veto an illusion by $\br$ only but leave behind an ambiguous percept 
 (like the invisibility of the reversed depth illusion in central vision in Fig. \ref{fig4}), 
in  Fig. \ref{fig5}, this occurs for $(\br , \brp )$  marked by blue-colored $P(S=1| {\bf r, r'})$.
To resolve the remaining perceptual ambiguity under $({\bf r, r'})$, another feedback query for 
some additional upstream responses is likely useful.  
For every  query, let ${\bf r}$ denote the whole collection of signals 
admitted after the previous queries (if any) and the initial feedforward signals,  
and $\brp$ the target of the current query, then, analogous to equation (\ref{eq:Get_br}), 
\begin{equation}
\begin{array}{l}
   \text{given available signal $\br$, and  a limited capacity (bottleneck) for querying additional signals, } \\ 
	\mbox{the target signal $\brp$ of the current feedback query should maximize $I(S; (\br, \brp))$.} \label{eq:Get_bR}
\end{array}
\end{equation}

The task set should be designed to choose $\br$ and $\brp$ optimally for the task, 
which is simply defined here by the $S$ to be decoded. 
Since the optimal target choice of $\br$ and $\brp$ depend on $S$,
the visual system requires some flexibility. 
The flexibility should be limited. For example, the feedforward and feedback neural circuits between
the brain areas along the visual pathway, and the recurrent 
neural circuit within each brain area, all impose limits.
Through quantities $H(S)$, $I(S; (\br, \brp ))$, 
and  $P(S | \br, \brp )$, the task set can 
include a metacognitive evaluation of the quality of decoding
by assigning a confidence in the perceived $\hat S$\citep{Fleming2024} and/or to decide whether 
to execute another iteration of the FFVW algorithm. 

Peripheral vision provides a very useful window to examine how
limited is the flexibility for selecting the feedforward signals $\br$.
This is not only through studies of
visual illusions, but also through studies of visual crowding\citep{WhitneyLevi2011}.
Fig. \ref{fig1}C demonstrates that, in peripheral vision, 
the orientation of the central bar in a $3\times 3$ array of bars 
is more legible, i.e., better decoded, when the bar is more salient 
by virtue of having a larger orientation contrast against the surrounding bars.
More (moderate amount of) practices for seeing the less salient 
bar does not substantially improve its legibility, suggesting a limit to the flexible 
selection of $\br$, at least in peripheral vision. 
We understand through V1SH that V1's intracortical circuit makes the 
salient bar evoke a higher V1 response than non-salient bars. 
The better decoding of the more salient bar may arise passively 
from this higher response, since a higher response improves decoding quality by a better 
signal-to-noise\citep{ZhaopingBook2014}. It may also be
a preferential or default choice for $\br$ to be from 
the most responsive V1 neurons to a scene. Behaviorally, to see more clearly a 
briefly appearing object (a bar) away from the gaze position in a cluttered image, 
it helps to frame this object by a prominent box and let this box appear about
100 ms before this object appears, even when observers already 
have a full knowledge of this object's upcoming location\citep{NakayamaMackeben1989}.
This suggests that much of the control for selecting the 
target $\br$ is triggered by exogenous visual inputs.

The observations in the last paragraph suggest that much of the control 
for selecting the target $\br$ for seeing is reflexive, likely 
by  some autonomous neural circuit operations (perhaps those in
V1 and subcortical structures).  
This suggests a limited flexibility 
to select $\br$ optimally by equation (\ref{eq:Get_br}) for any 
given $S$. 
(One should however keep in mind that these observations
come from peripheral vision, which is specialized for looking rather than
seeing according to our framework. Hence, it could be that
there is a larger  flexibility for selecting the target $\br$ 
in central vision, which is specialized for seeing. Meanwhile, 
since central vision can employ the feedback query, flexibility for
$\br$ may be inessential.)  Instead, it is likely that, through evolution 
and learning, the neural circuit operations  adapt to maximize 
the average of  $I(S; \br )$ over the ensemble of tasks or $S$'s 
experienced in an animal's lifetime in its ecological habitat. 

The limit in controlling the selection of $\br$ based on the task or $S$
makes it important to achieve flexible control in selecting the target $\brp$ 
for feedback query based on $S$, according to equation  (\ref{eq:Get_bR}).  
Our VBC framework's non-trivial behavioral predictions, some of which enjoy 
experimental confirmation, can hopefully motivate investigations into the neural implementation
of  such top-down control, particularly in the FFVW algorithm. 
Traditional and ongoing research on top-down visual attention are most 
likely related to provide useful clues\citep{DesimoneDuncan1995,Carrasco2011}.
Meanwhile, according to the VBC framework, the top-down query mainly
targets the central visual field. Hence, one of the most 
effective forms of control is to shift gaze to another task relevant location,
such as to shifting from the eye region to the mouth region of 
a face image to better recognize emotion.  
Making a gaze shift not only alters the visual samples 
on the visual scene through a highly non-uniform cone 
density in the retina (and thus obtaining $\brp$ through 
the new retinal samples), but also enables a better control over 
selecting $\brp$ at an otherwise peripheral visual location without this gaze shift.
By the VBC perspective, many small-amplitude gaze shifts (within one degree of
visual angle) during a gaze fixation, often called fixational eye movements, 
are also likely for  selecting $\brp$ rather than reflecting random noise in (and the control of)
gaze position\citep{RucciPoletti2015}.

%Meanwhile, a certain degree of endogenous control is
%suggested by the following observation. When a monkey has to saccade 
%to an object that differs from seven other identical objects, 
%all equal distant from the initial gaze position, 
%a lesion in V4 downstream from V1 along the visual pathway 
%particularly impairs the task performance when the odd-ball target
%is less prominent (e.g., dimmer or smaller) than the other 
%objects\citep{SchillerLee1991}.

\section{Discussion and Summary}

The VBC framework is not only motivated by the need to place the critically overlooked 
attentional bottleneck at the center stage of vision, but is also highly unusual in 
the modern field of vision science by being a theoretical framework that provides  
non-trivial and easily falsifiable predictions. Some of which are subsequently 
confirmed (Figs. \ref{fig2}, \ref{fig3}, \ref{fig4}).
Such easily testable predictions can hopefully serve as stepping
stones to move the field beyond the decades-old frontier 
separating the better understood V1 from the less understood visual 
stages downstream from V1. 

For example, the prediction that the top-down feedback to V1 from downstream stages 
involved in recognition (in the ventral stream of the visual pathway, see
Fig. \ref{fig1}B) should be mainly directed to the central visual field 
can be easily tested through anatomical experiments. 
Another prediction is that the proportion of simple cells 
(versus complex cells) in V1 should be higher among 
neurons covering the central rather than the peripheral visual field.
This is because simple cells are more sensitive than complex cells 
to the precise spatial locations of visual features to better serve feedback 
queries for additional details of visual inputs.
This prediction may also apply to V2 and higher early visual stages
along the visual pathway, particularly if the bottleneck starting from 
V1 to downstream areas is gradual so that the feedback query could
also query from (e.g.,) V2 responses.  
This prediction can also be easily tested through electrophysiological 
experiments. Previous experimental investigations have not examined 
how such neural properties, in the prevalence of the feedback fibers or 
in the proportion of simple versus complex cells, depend on the eccentricity of  
visual field locations, since there had been no theoretical motivation
for such investigations.  

Given the reversed feature illusions, e.g., the flip tilt illusion, the
reversed motion illusion, and the reversed depth illusion, 
one can examine the presence or absence of neural responses 
to the reverse feature signals in visual areas downstream of
V1.  Whether and how neurons respond to these reversed feature signals in each
downstream cortical area, and how such neural responses 
depend on the eccentricity of the visual field locations, 
and how such responses vary in time relative to the stimulus onset, 
should shed light on the neural implementation of the FFVW and 
FfW algorithms.

V1 is one of the two largest visual cortical areas, occupying 21\% of total cortical 
area devoted to vision in a primate brain\citep{FellemanVanEssen1991}. 
Since the elucidation of V1 neural receptive fields in early 1960s\citep{HubelWiesel1962},
it took a few decades before we discovered V1's role in looking\citep{LiPNAS1999, LiTICS2002}.
V2, immediately downstream from V1, has about the same size as V1.  
Its neurons appear similar to V1 neurons in their selectivity to visual features, while 
their neural receptive fields are about 2-3 times as large as those of 
the V1 neurons\citep{GattassEtAl1981, FosterEtAl1985}. 
A challenge to the VBC framework is that it should help us discover and understand 
V2's role in vision.  With the VBC  proposal that the bottleneck starts from 
V1's output to V2, and since seeing (visual inference or decoding) should occur after 
the bottleneck, I  propose that V2 further deletes visual information by decoding 
or making inferences about global properties of object 
surfaces in 3D space from image features in visual input images. 
From an informatics perspective, this inference serves to quantize visual information in 
2-dimensional images into coarser categories based on properties of the 
underlying object surfaces in 3D space, thereby deleting visual image information 
less relevant for representing global shape and statistical characters of the surfaces.
Perceptually, this is likely related to such psychological phenomena  as
surface completion, figure-ground segregation, contour integration, 
and various forms of Gestalt grouping.  Neurally, this should be linked 
with the observed V2 neural properties including sensitivity  to 
illusory contours\citep{VonderHeydtEtAl1984},  border-ownership\citep{ZhouFriedmanHeydt2000}, 
stereoscopic edges between surfaces\citep{VonderHeydtEtAl2000}, sensitivity to relative 
depth between depth surfaces\citep{ThomasEtAl2002},  and sensitivity to 
surface configurations\citep{BakinEtAl2000}. Computationally, inference of 
object surfaces is a stepping stone (cf. mid-level vision) towards visual 
segmentation and object recognition. Both intra-V2 recurrent circuits and interactions 
between V1, V2 and other downstream areas are likely to play significant roles.  
Indeed, neural tuning to depth edges and border ownerships
have been shown to emerge in models of V2 with short-range intra-V2 interactions 
between model V2 neurons\citep{ZhaopingStereo2002,ZhaopingNeuron2005}.
Understanding V2 should also help us to answer the following question: how much  
of the visual information loss by  the bottleneck occurs between V1 and V2?

In summary, the VBC framework offers a new path to understanding
vision by placing the bottleneck in the center stage and by formulating vision
as mainly looking and seeing through the bottleneck.
Since computational resources are limited in all biological brains to exert a bottleneck, 
and since limited resources  are shared across multiple sensory systems 
(typically including visual, auditory,  somatosensory,  and olfactory systems), 
the central-peripheral dichotomy should generalize across animal species and also generalize 
multisensorily, so that looking and seeing should generalize to sensory orienting and sensory inferencing\citep{ZhaopingCPD2023}. 
Hence, for example, the peripheral visual field in primates should extended to 
include the auditory and other sensory fields. Meanwhile, the central sensory field is mainly by, 
for example, visual fovea in primates, olfactory and somatosensory fovea in rodents, 
and auditory fovea in bats.  Through evolution, the neural hardware for reflexive 
guidence to orienting evolve from the optic tectum (called superior colliculus in mammals) 
in lower vertebrates to V1 in primates\citep{Zhaoping2016Evolution}. 
Meanwhile, there should be no clear or definite boundary between ``seeing" and 
understanding.  At least in primates, the computation of analysis-by-synthesis, 
using the FFVW algorithm, requires knowledge linking the 3D visual world to 2D visual 
images and their neural responses. Anatomically, the higher visual cortical areas 
such as IT and the frontal eye fields (Fig. \ref{fig1}B) 
are closer to frontal brain areas for executive functions.  
Studying vision is truely looking into our brain through our eyes, 
while every animal species offers a window to understanding cognition and  intelligence 
across  species through our search for unifying principles.

\bibliography{ReferenceForBook}

\end{document}